\documentclass[preprint,showpacs,preprintnumbers,amsmath,amssymb]{revtex4}
 \usepackage{amsmath}
 \usepackage{txfonts}
 \usepackage{amssymb}
 \usepackage{graphicx,epsfig}
 \usepackage{dcolumn}
 \usepackage{bm}
 \begin{document}

 \title{Topological aspects in a two component
Bose condensed system in neutron star}
  \author{Ji-Rong Ren}
  \author{Heng Guo}\email{guoh06@lzu.cn}
 \affiliation{Institute of Theoretical Physics, Lanzhou
 University, Lanzhou 730000, P. R. China}

 \begin{abstract}

By making use of Duan-Ge's decomposition theory of gauge potential
  and the Duan's topological current theory proposed by Prof. Duan Yi-Shi,
  we study a two component superfluid Bose condensed system, which is supposed being
 realized in the interior of neutron stars in the form of a
 coexistent neutron superfluid and protonic superconductor. We
 propose that this system possesses vortex lines. The topological
 charge of the vortex lines are characterized by the Hopf indices
 and the Brower degrees of $\phi$-mapping.
 \end{abstract}

 %\\Key words: {Neutron stars, Topological excitations, Spontaneous breaking of gauge symmetries}

 \maketitle
\section{Introduction}

In the standard model for a neutron star its interior features
superfluidity of neutron Cooper pairs and superconductivity of
proton Cooper pairs (see,e.g.,\cite{Mhoffberg,Gbaym}). Both these
condensates allow vortices of the $S^{1}\rightarrow S^{1}$ map. It
was suggested that the phenomenon of glitches in Crab and Vela
pulsar is connected with vortex matter in these stars
\cite{vortexinstars}. The neutron star spin down can be explained by
neutrino radiation from superfluid vortex \cite{peng}. This remains
a topic of intensive studies and discussions (for recent
developments and citations see \cite{blink}\cite{zuo}). Besides that
a standard model for a neutron star is a special system being a two
component superfluid Bose condensed system which makes it also being
a topic of abstract academic interest \cite{bcarter} since such a
system allows for interesting phenomena with no direct counterparts
in e.g. superconducting metals. Since the two component superfluid
Bose condensed system have played an important roles in the
condensed matter and neutron stars, studies of topological defects
in this system becomes an important study aspect \cite{babaev}.

In this paper we mainly focus attention on the two component
superfluid Bose condensed system, which is believed to be realized
in the interior of neutron stars. We start by the free energy
density of the two component superfluid Bose condensed system. By
making use of Duan-Ge's decomposition theory of gauge potential and
the Duan's topological current theory proposed by Prof. Duan Yi-Shi
\cite{duan1}, we propose that this system possesses vortex lines,
and the topological charge of the vortex lines are characterized by
the Hopf indices and the Brower degrees of $\phi$-mapping.

\section{Two component Bose condensed system}

We consider the following effective Landau-Ginzburg free energy
density that describes a two component superfluid Bose condensed
system. In this system, we have a proton condensate described by
$\psi_{1}$ and a neutron condensate described by $\psi_{2}$. We do
not consider the normal component of the protons and neutrons with
their specific excitations, only the superfluid component. The
$\psi_{1}$ field with electric charge q, which is actually twice the
fundamental charge of the proton $q=2|e|$, interacts with the gauge
field $\vec{A}$, with $\vec{B}=\nabla\times\vec{A}$. The free energy
density reads the following expression\cite{prl92}:
\begin{eqnarray}\label{fed}
f &=& \frac{\hbar^{2}}{2M_{1}}\big|(\nabla -\frac{iq}
       {\hbar c}\vec{A})\psi_{1}\big|^{2}\nonumber\\
  &+& \frac{\hbar^{2}}{2M_{2}}\big|\nabla\psi_{2}\big|^{2}
     + \frac{\vec{B}^{2}}{8\pi} + V(|\psi_{1}^{2}|,|\psi_{2}^{2}|),
\end{eqnarray}
where $M_{1}=2m_{1}$ and $m_{1}$ is the mass of the proton, and
$M_{2}=2m_{2}$ and $m_{2}$ is the mass of the neutron. In the free
energy density given above, we have ignored the term coupling the
proton and neutron superfluid velocities, which gives rise to the
Andreev-Bashkin effect\cite{JETP42}, as it is not important in our
discussion. Indeed, the relevant term in the free energy can be
represented as $\sim\int d^{3}x\vec{v}_{1}\cdot\vec{v}_{2}$, where
$\vec{v}_{1}$ and $\vec{v}_{2}$ are velocities of the superfluid
components. For neutron stars which do not rotate (the case which is
considered in this paper), $\vec{v}_{2}=0$, and the effect obviously
vanishes. We expect that due to the small density of the neutron
vortices (compared to the density of the proton vortices) the effect
is still negligible for most of the flux tubes in a rotating star as
well \cite{prl92}. The effect could be important only for a few of
the flux tubes situated close to a neutron vortex core, where
$\vec{v}_{2}$ strongly deviates from the constant value at interflux
distance scales. We have also ignored the fact that the neutron
condensate has a nontrivial $^{3}p_{2}$ order parameter as only the
magnitude of the neutron condensate is relevant to the effect
described below. The free energy density (\ref{fed}) only describes
large distances and it does not describe the gap structure on the
Fermi surfaces, only the superfluid component of the protons and
neutrons.

For discussing the detail of Eq. (\ref{fed}), we start by
introducing variables $\chi_{1}$, $\chi_{2}$ and $\rho$ by
\begin{eqnarray}\label{psi1and2}
\psi_{1}=\sqrt{2M_{1}}\rho\chi_{1}, \quad\quad
\psi_{2}=\sqrt{2M_{2}}\rho\chi_{2}
\end{eqnarray}
where the complex $\chi_{1}=|\chi_{1}|e^{i\phi_{1}}$ and
$\chi_{2}=|\chi_{2}|e^{i\phi_{2}}$ are chosen so that
$|\chi_{1}|^{2}+|\chi_{2}|^{2}=1$. The modulus $\rho$ then has the
following expression:
\begin{eqnarray}\label{rho}
\rho^{2}=\frac{1}{2}[\frac{|\psi_{1}|^{2}}{M_{1}}+\frac{|\psi_{2}|^{2}}{M_{2}}].
\end{eqnarray}
By putting formula (\ref{psi1and2}) to use, the free energy density
(\ref{fed}) can be represented as
\begin{eqnarray}\label{fed2}
f &=& \hbar^{2}(\nabla\rho)^{2}
    + \hbar^{2}\rho^{2}\big|(\nabla-\frac{iq}{\hbar
        c}\vec{A})\chi_{1}|^{2}
    + \hbar^{2}\rho^{2}\big | \nabla\chi_{2}|^{2}\nonumber\\
  &+& \frac{\vec{B}^{2}}{8\pi} + V.
\end{eqnarray}
Then we can introduce current density
\begin{eqnarray}\label{current}
\vec{J}&=&\frac{i\hbar}{2M_{1}}(\psi_{1}\nabla\psi_{1}^{*}-\psi_{1}^{*}\nabla\psi_{1})
          + \frac{i\hbar}{2M_{2}}(\psi_{2}\nabla\psi_{2}^{*}-\psi_{2}^{*}\nabla\psi_{2})
           \nonumber\\
        &-&\frac{q}{M_{1}c}\vec{A}|\psi_{1}|^{2}
\end{eqnarray}
and electrical current density
\begin{eqnarray}\label{ecurrent}
\vec{J_{q}}=\frac{i\hbar q}{2M_{1}}
              (\psi_{1}\nabla\psi_{1}^{*}-\psi_{1}^{*}\nabla\psi_{1})
              -\frac{q^{2}}{M_{1}c}\vec{A}|\psi_{1}|^{2}
\end{eqnarray}
Then we can get
\begin{eqnarray}\label{current2}
\vec{J}= \hbar\rho^{2}\vec{j}
          -\frac{2\rho^{2}q}{c}\vec{A}|\chi_{1}|^{2},
\end{eqnarray}
where
\begin{eqnarray}\label{jcurrent}
\vec{j}= i(\chi_{1}\nabla\chi_{1}^{*}-\chi_{1}^{*}\nabla\chi_{1}
           + \chi_{2}\nabla\chi_{2}^{*}-\chi_{2}^{*}\nabla\chi_{2}).
\end{eqnarray}
We introduce a gauge-invariant vector field $\vec{C}$, directly
related to the current density by
\begin{eqnarray}\label{c_vec}
\vec{C}=\frac{\vec{J}}{\rho^{2}}.
\end{eqnarray}
We can also get the electrical current density
\begin{eqnarray}\label{ecurrent1}
\vec{J_{q}}= \hbar q\rho^{2}[i(\chi_{1}\nabla\chi_{1}^{*}
                 -\chi_{1}^{*}\nabla\chi_{1})-
                  \frac{2q}{c\hbar}\vec{A}|\chi_{1}|^{2}].
\end{eqnarray}

We then rearrange the terms in Eq. (\ref{fed2}) as follows: we add
and subtract from Eq. (\ref{fed2}) a term
$\frac{1}{4}\hbar^{2}\rho^{2}\vec{j}$ and observe that the following
expression:
\begin{eqnarray}\label{term}
\hbar^{2}\rho^{2}\big[|\nabla\chi_{1}|^{2} + |\nabla\chi_{2}|^{2}
                    - \frac{\vec{j}}{4}^{2}\big]
\end{eqnarray}
is also gauge invariant. Indeed if we introduce the gauge invariant
field
\begin{eqnarray}\label{n_vec}
\vec{n}=\overline{\chi}\vec{\sigma}\chi,
\end{eqnarray}
where $\overline{\chi}=(\chi_{1}^{*}, \chi_{2}^{*})$ and
$\vec{\sigma}$ are Pauli matrices, then $\vec{n}$ is a unit vector
$|\vec{n}|=1$ and we can write Eq. (\ref{term}) as follows:
\begin{eqnarray}\label{term2}
\hbar^{2}\rho^{2}\big[|\nabla\chi_{1}|^{2} + |\nabla\chi_{2}|^{2}
                    - \frac{\vec{j}}{4}^{2}\big]
 = \frac{1}{4}\hbar^{2}\rho^{2}(\nabla\vec{n})^{2}.
\end{eqnarray}

Now we consider the remaining terms in Eq. (\ref{fed2}). For the
magnetic field we can get
\begin{eqnarray}\label{magnetic}
\vec{B} = \nabla\times\vec{A}
        = \frac{\hbar c}{2q|\chi_{1}|^{2}}
           (\nabla\times\vec{j}-\frac{1}{\hbar}\nabla\times\vec{C})
\end{eqnarray}
where $\nabla\times\vec{j}$ can be rewritten by the unit vector
$\vec{n}$ as following expression:
\begin{eqnarray}\label{rotj}
\nabla\times\vec{j}=\frac{1}{2}\vec{n}\cdot
    \partial_{\mu}\vec{n}\times\partial_{\nu}\vec{n}.
\end{eqnarray}
Combining these we can obtain our main result: the free energy
density becomes
\begin{eqnarray}\label{fedresult}
f &=& \hbar^{2}(\nabla\rho)^{2}
      + \frac{1}{4}\hbar^{2}\rho^{2}(\nabla\vec{n})^{2} \nonumber\\
  &+& \frac{\hbar^{2}c^{2}}{128\pi q^{2}|\chi_{1}|^{4}}
         \big[\frac{1}{\hbar}(\partial_{\mu}C_{\nu}-\partial_{\nu}C_{\mu})
           - \vec{n}\cdot\partial_{\mu}\vec{n}\times\partial_{\nu}\vec{n}\big]^{2}
           \nonumber\\
 &-&  \frac{\hbar}{2q|\chi_{1}|^{2}}\vec{J_{q}}\cdot(\vec{j}-\frac{\vec{C}}{\hbar})
      + \frac{1}{4}\hbar^{2}\rho^{2}\big[\vec{j}\;^{2}
      - \frac{1}{|\chi_{1}|^{2}}(\vec{j}-\frac{\vec{C}}{\hbar})^{2} \big]\nonumber\\
 &+& V(\rho, |\chi_{1}|^{2}, |\chi_{2}|^{2}).
\end{eqnarray}
Let us discuss topological defects, allowed in Eq.
(\ref{fedresult}).

\section{Topological vortex lines in the two component Bose condensed system}

As shown in Eq. (\ref{fedresult}), we know that the magnetic field
of the system can be divided into two parts: One is the contribution
of field $C_{\mu}$, we learn that this part is introduced by the
current density and can only present us with the topological defects
as what in a single-condensate system. The other part, the
contribution
$\vec{n}\cdot\partial_{\mu}\vec{n}\times\partial_{\nu}\vec{n}$ to
the magnetic field term in Eq. (\ref{fedresult}), is a fundamentally
topological property of this system. Here we emphasize that there
allow another nontrivial topological configurations, which are
originated from the contribution of the term
$\vec{n}\cdot\partial_{\mu}\vec{n}\times\partial_{\nu}\vec{n}$. Thus
we will investigate this term in detail.

It is easy to prove that this term
$\vec{n}\cdot\partial_{\mu}\vec{n}\times\partial_{\nu}\vec{n}$ can
be reexpressed in an Abelian field tensor form
\begin{equation}\label{tensor}
    \vec{n}\cdot\partial_{\mu}\vec{n}\times\partial_{\nu}\vec{n}
     = \partial_{\mu}b_{\nu}-\partial_{\nu}b_{\mu}
\end{equation}
where $b_{\mu}$ is the Wu-Yang potential \cite{ttwu}
\begin{equation}\label{wuyang}
    b_{\mu}=\vec{e_{1}}\cdot\partial_{\mu}\vec{e_{2}}
\end{equation}

Here $\vec{e_{1}}$ and $\vec{e_{2}}$ are two perpendicular unit
vectors normal to $\vec{n}$, and $(\vec{e_{1}}, \vec{e_{2}},
\vec{n})$ forms an orthogonal frame:
\begin{eqnarray}\label{unitvec}
\vec{e_{1}}\cdot\vec{e_{2}}=\vec{e_{1}}\cdot\vec{n}=\vec{e_{2}}\cdot\vec{n}=0,\nonumber\\
    \vec{e_{1}}\cdot\vec{e_{1}}=\vec{e_{2}}\cdot\vec{e_{2}}=\vec{n}\cdot\vec{n}=1
\end{eqnarray}

Now, consider a two-component vector field $\vec{\phi}=(\phi^{1},
\phi^{2})$ residing in the plane formed by $\vec{e_{1}}$ and
$\vec{e_{2}}$:
\begin{equation}\label{phifield}
    e_{1}^{a}=\frac{\phi^{a}}{||\phi||} \quad
    e_{2}^{a}=\epsilon^{ab}\frac{\phi^{b}}{||\phi||} \quad
    (||\phi||^{2}=\phi^{a}\phi^{a}; a, b=1, 2)
\end{equation}

It can be proved that relations for $\vec{e_{1}}$ and $\vec{e_{2}}$
satisfy above restriction (\ref{unitvec}). Obviously the zero point
of $\vec{\phi}$ are just the two-dimensional singular points of
$\vec{e_{1}}$ and $\vec{e_{2}}$.

Using the $\vec{\phi}$ field, the Wu-Yang potential can be expressed
as
\begin{equation}\label{wy2}
    b_{\mu}=\epsilon_{ab}\frac{\phi^{a}}{||\phi||}\partial_{\mu}\frac{\phi^{b}}{||\phi||}.
\end{equation}

Comparing with the expression of Duan-Ge's decomposition theory of
gauge potential \cite{duan2}, we learn that $b_{\mu}$ satisfies the
$U(1)$ gauge transformation. We introduce a two order tensor
\begin{equation}\label{tensorh}
    H_{\mu\nu}=\vec{n}\cdot\partial_{\mu}\vec{n}\times\partial_{\nu}\vec{n}
\end{equation}
which describes the magnetic field that becomes induced in the
system due to a non-trivial electranagnetic interaction in this
system.

By using the $\vec{\phi}$ field, the two order tensor $H_{\mu\nu}$
can be reexpressed as
\begin{equation}\label{tensorh2}
    H_{\mu\nu}=2\epsilon_{ab}\partial_{\mu}\frac{\phi^{a}}{||\phi||}\partial_{\nu}\frac{\phi^{b}}{||\phi||}
\end{equation}

Because this tensor $H_{\mu\nu}$ plays an important role in the
topological feature in this system, here we will using the the
Duan's topological current theory, to research the properties hidden
in this tensor. To do so, we introduce a topological tensor current
$K^{\mu\nu}$ \cite{duan2}, which is denoted by
\begin{equation}\label{topocurrent}
    K^{\mu\nu}=\frac{1}{8\pi}\epsilon^{\mu\nu\lambda\rho}H_{\lambda\rho}
              =\frac{1}{4\pi}\epsilon^{\mu\nu\lambda\rho}
              \epsilon_{ab}\partial_{\mu}\frac{\phi^{a}}{||\phi||}\partial_{\nu}\frac{\phi^{b}}{||\phi||}.
\end{equation}

According to \cite{duan1} and using
$\partial_{\mu}\frac{\phi^{a}}{||\phi||}=\frac{\partial_{\mu}\phi^{a}}{||\phi||}
+ \phi^{a}\partial_{\mu}\frac{1}{||\phi||}$ and the Green function
relation in $\phi$-space:
\begin{equation}\label{greenf}
    \partial_{a}\partial_{a}ln||\phi||=2\pi\delta^{2}(\vec{\phi})
    \quad\quad
    (\partial_{a}=\frac{\partial}{\partial\phi^{a}}),
\end{equation}
it can be proved that
\begin{equation}\label{topocurrentk}
    K^{\mu\nu}=\delta^{2}(\vec{\phi})D^{\mu\nu}(\frac{\phi}{x})
\end{equation}
where $D^{\mu\nu}(\frac{\phi}{x})=
\frac{1}{2}\epsilon^{\mu\nu\lambda\rho}\epsilon_{ab}\partial_{\lambda}\phi^{a}\partial_{\rho}\phi^{b}$
is the Jacobian between the $\phi$ space and the Cartesian
coordinates. Denoting the spacial component of $K^{\mu\nu}$ by
$K^{i}$, we obtain
\begin{equation}\label{ki1}
    K^{i}=K^{0i}=\delta^{2}(\vec{\phi})D^{i}(\frac{\phi}{x})
    \quad\quad (i=1, 2, 3)
\end{equation}
where $D^{i}(\frac{\phi}{x})=D^{0i}(\frac{\phi}{x})$.

An important conclusion form formula (\ref{ki1}) is
\begin{equation} \label{ki2}
\left\{
 \begin{aligned}
     K^{i} &=& 0 \quad\quad \mbox{if and only if}\quad \vec{\phi}\neq 0\\
     K^{i} &\neq& 0 \quad\quad \mbox{if and only if}\quad
     \vec{\phi}=0.
 \end{aligned} \right.
\end{equation}

So it is necessary to study the zero point of $\phi$ to determine
the nonzero solution of $K^{i}$. The implicit function theory
\cite{goursat} show that under the regular condition
$D^{i}(\frac{\phi}{x})\neq 0$, the general solution of
\begin{equation}\label{phi12}
    \phi^{1}(t,x^{1},x^{2},x^{3})=0, \quad \quad
    \phi^{1}(t,x^{1},x^{2},x^{3})=0
\end{equation}
Can be expressed as
\begin{eqnarray}\label{xk}
x^{1}&=& x_{k}^{1}(s,t)\quad\quad x^{2}=x_{k}^{2}(s,t)
 \quad\quad\nonumber\\
  x^{3}&=& x_{k}^{3}(s,t) \quad\quad (k=1,2,3,\cdots,N)
\end{eqnarray}
which represent the world surfaces of $N$ moving isolated singular
strings $L_{k}$ with string parameter $s$. This indicates that in
this system, there are vortex lines located at the zero points of
the $\vec{\phi}$ field.

Now we will study the topological charges of the vortex lines and
their properties by making using of the Duan's topological current
theory. In $\delta$-function theory\cite{jaschouten}, one can prove
that in three-dimensional space
\begin{equation}\label{delta2}
    \delta^{2}(\vec{\phi})=\sum_{k=1}^{N}\beta_{k}\int_{L_{k}}
    \frac{\delta^{3}(\vec{x}-\vec{x_{k}}(s))}{|D(\frac{\phi}{u})|_{\sum_{k}}}
\end{equation}
where
\begin{equation}\label{dfunction}
    D\big(\frac{\phi}{u}\big)_{\sum_{k}}=\frac{1}{2}\epsilon^{\mu\nu}\epsilon_{mn}
    \big(\frac{\partial\phi^{m}}{\partial u^{\mu}}\big)\big(\frac{\partial\phi^{n}}{\partial u^{\nu}}\big)
\end{equation}
and $\sum_{k}$ is the $k$th planar element transverse to $L_{k}$
with local coordinates $(u^{1},u^{2})$. The positive integer
$\beta_{k}$ is the Hopf index of $\phi$-mapping, which means that
when the point $\vec{x}$ covers the neighborhood of the zero point
$\vec{x}_{k}$ once. The vector field $\vec{\phi}$ covers the
corresponding region in $\phi$-space $\beta_{k}$ times. Meanwhile,
the direction vector of $L_{k}$ is given by \cite{duan1}
\begin{equation}\label{dxds}
    \left. \frac{dx^{i}}{ds}\right|_{\vec{x}_{k}}=
     \left.\frac{D^{i}(\phi/x)}{D(\phi/u)}\right|_{\vec{x}_{k}},
\end{equation}
which leads to
\begin{equation}\label{dxdx}
\left. \frac{dx^{1}}{dx^{3}}\right|_{\vec{x}_{k}}=
     \left. \frac{D^{1}(\phi/x)}{D^{3}(\phi/x)}\right|_{\vec{x}_{k}},
     \quad
\left. \frac{dx^{2}}{dx^{3}}\right|_{\vec{x}_{k}}=
     \left. \frac{D^{2}(\phi/x)}{D^{3}(\phi/x)}\right|_{\vec{x}_{k}}.
\end{equation}
There from (\ref{delta2}) and (\ref{dxds}). We find the inner
topological structure of
\begin{equation}\label{ki}
K^{i}=\delta^{2}(\vec{\phi})D^{i}(\phi/x)
      =\sum_{k=1}^{N}\beta_{k}\eta_{k}\int_{L_{k}}
         \frac{dx^{i}}{ds}\delta^{3}(\vec{x}-\vec{x}_{k}(s))ds
\end{equation}
in which $\eta_{k}$ is the Brouwer degree of the $\phi$-mapping,
with
\begin{equation}\label{etak}
    \eta_{k}=sgn D(\phi/u)_{\sum_{k}}=\pm 1.
\end{equation}
Hence the topological charge of the vortex line $L_{k}$ is
\begin{equation}\label{Qk}
    Q_{k}=\int_{\sum_{k}}K^{i}d\sigma_{i}=W_{k},
\end{equation}
where $W_{k}=\beta_{k}\eta_{k}$ is the winding number of
$\vec{\phi}$ around $L_{k}$. And the total topological number on
surface $\sum$ is
\begin{equation}\label{Q}
    Q=\int_{\sum}K^{i}d\sigma_{i}=\sum_{k=1}^{n}W_{k}.
\end{equation}
Then we come the conclusion: under the regular condition
$D^{i}(\phi/x)\neq 0$ there exist vortex lines in this system, whose
topological charge are just the winding numbers of the
$\phi$-mapping.

\section{Conclusion}
In conclusion, based on the Duan-Ge's decomposition theory of gauge
potential and the Duan's topological current theory, we point out
there allow vortex lines in the two component superfluid Bose
condensed system, which is believed to be realized in the interior
of neutron stars. Under the regular condition $D^i(\phi/x)\neq0$,
vortex lines are originated form the zero points of the $\vec\phi$
field and the topological charges of vortex lines are characterized
by the winding numbers of the $\phi$-mapping.

In this paper we treat the vortex lines as mathematical lines, i.e.,
the width of a line is zero. This description is obtained in the
approximation that the radius of curvatures of a line is much larger
than the width of the line \cite{nielsen}.

At last, it should be pointed out that in the present paper, we just
considered the case in which the neutron star does not rotate, but
the effect of the neutron star's rotation is significant. We will
study the topological aspects of rotary neutron star in our next
work.

\section{Acknowledgement}

One of the authors Guo H. would like to thank Dr. Xu Dong-Hui and
Dr. Zhang Xin-Hui for numerous fruitful discussions and help. This
work was supported by National Natural Science Foundation of P. R.
China.


\begin{thebibliography}{99}

\bibitem{Mhoffberg}
M. Hoffberg \emph{et al.},phys. Rev. Lett. \textbf{24}, 175 (1970).

\bibitem{Gbaym}
G. Baym, C. Pethick, and D. Pines, Nature (London) \textbf{224}, 673
(1969).

\bibitem{vortexinstars}
P.W. Anderson, N. Itoh, Nature \textbf{256} 25 (1975); M. A. Alpar,
Astroph. J. \textbf{213} 527 (1977); P.W. Anderson \emph{et. al.}
Phil. Mag. A \textbf{45} 227 (1982); D. Pines \emph{et. al.} Progr.
Teor. Phys. Suppl. \textbf{69} 376 (1980); M.A. Alpar \emph{et. al.}
Astroph. J. (Letters) \textbf{249} L33 (1982)

\bibitem{peng}
Q.H.Peng, K.L.Huang, J.H.Huang, Astron\&Astrophys, \textbf{107} 258
(1982)

\bibitem{blink}
B. Link Phys.Rev.Lett. \textbf{91} 101101 (2003); K. B.W. Buckley,
M. A. Metlitski, A. R. Zhitnitsky, Phys.Rev. C \textbf{69} 055803
(2004)

\bibitem{zuo}
Cui Chang-Xi, Zuo Wei, and H. J. Schulze, Chin.Phys.B \textbf{17}
3289 (2008), Li Z. H., Li Z. H., Zuo W., and Luo P. Y.  Acta. Phys.
Sin. \textbf{55} 84 (2006) (in Chinese)

\bibitem{bcarter}
For recent papers and citations see B. Carter, astro-ph/0101257; D.
Langlois, astro-ph/0008161, B. Carter and D. Langlois, Nucl. Phys. B
\textbf{531}, 478 (1998)

\bibitem{babaev}
E. Babaev, L. D. Faddeev, and A. J. Niemi,  Phys. Rev. B
\textbf{65}, 100512(R) (2002).


\bibitem{duan1}
Y. S. Duan, H. Zhang and S. Li, Phys. Rev. B \textbf{58} 125 (1998);
Y.S. Duan, S. Li, and G.H. Yang, Nucl. Phys. B \textbf{514}, 705
(1998); Y. Jiang and Y.S. Duan, J. Math. Phys. \textbf{24}, 6463
(2000). Y. S. Duan, S. L. Zhang \emph{et al.}, J. Math. Phys.
\textbf{35} 4463 (1994)

\bibitem{prl92}
K. B. W. Buckley, M. A. Metlitski, and A. R. Zhitnitsky, Phys. Rev.
Lett. \textbf{92} 151102 (2004).

\bibitem{JETP42}
A. F. Andreev and E. Bashkin, Sov. Phys. JETP \textbf{42}, 164
(1975)

\bibitem{ttwu} T.T. Wu, and C.N. Yang, Phys. Rev. D \textbf{12}, 3845 (1975); \textbf{14}, 437
(1975).

\bibitem{duan2} Y.S. Duan, X. Liu, and P.M. Zhang, J. Phys.
A \textbf{36}, 563 (2003); Y.S. Duan, H. Zhang, and S. Li, Phys.
Rev. B \textbf{58}, 125 (1998);  Y.S. Duan, X.H. Zhang, Y.X. Liu and
L. Zhao, Phys. Rev. B \textbf{74}, 144508 (2006).

\bibitem{goursat}
\`{E}. Goursat, \emph{A Course in Mathematical Analysis}, translated
by E. R. Hedrick (Dover, New York, 1904), Vol.I.

\bibitem{jaschouten}
J. A. Schouten, \emph{Tensor Analysis for Physics} (Clarendon,
Oxford, 1951)

\bibitem{nielsen} H.B. Nielsen and P. Olesen, Nucl. Phys. B \textbf{61}, 45
(1973).



\end{thebibliography}
\end{document}